\documentclass[sn-mathphys-num]{sn-jnl}

\usepackage[T1]{fontenc}
\usepackage{amsmath}
\usepackage{amsfonts}
\usepackage{graphicx}
\usepackage{hyperref}
\usepackage{url}
\usepackage{doi}

\makeatletter
\def\doiurl#1{\begingroup\urlstyle{tt}\url{#1}\endgroup}
\makeatother

\usepackage{multirow}
\usepackage{longtable}
\usepackage{array}
\usepackage{siunitx}
\usepackage{mathrsfs}
\usepackage[title]{appendix}
\usepackage{xcolor}
\usepackage{textcomp}
\usepackage{manyfoot}
\usepackage{booktabs}
\usepackage{algorithm}
\usepackage{algorithmicx}
\usepackage{algpseudocode}
\usepackage{listings}

\title[The Face of Populism]{The Face of Populism: Examining Differences in Facial Emotional Expressions of Political Leaders Using Machine Learning}

\author[1]{\fnm{Sara} \sur{Major} }\email{sara.mejdzor@ff.uns.ac.rs}
\author*[1]{\fnm{Aleksandar} \sur{Tomašević}}\email{atomashevic@ff.uns.ac.rs}

\affil*[1]{\orgdiv{Department of Sociology}, \orgname{University of Novi Sad}, \orgaddress{\city{Novi Sad}, \country{Serbia}}}

\begin{document}

\maketitle

\begin{center}
    
{\color{blue} Please refer to and cite the published version: } \\
\vspace{0.2cm}
Major, S., Tomašević, A. The face of populism: examining differences in facial emotional expressions of political leaders using machine learning. J Comput Soc Sc 8, 62 (2025). \url{https://doi.org/10.1007/s42001-025-00392-w}
\end{center}
\section*{Abstract}

\begin{abstract}

Populist rhetoric employed in online media is characterized as deeply impassioned and most often imbued with strong emotions. This paper investigates the differences in affective non-verbal communication of political leaders. We use a deep-learning approach to process a sample of 220 YouTube videos depicting political leaders from 15 different countries, analyze their facial expressions of emotion, and then examine differences in average emotion scores representing the relative presence of 6 emotional states (anger, disgust, fear, happiness, sadness, and surprise) and a neutral expression for each frame of the processed YouTube video. Based on a sample of manually coded images, we find that this machine learning approach has 53-60\% agreement with human annotation. We observe statistically significant differences in the average score of expressed negative emotions between groups of leaders with varying degrees of populist rhetoric. Overall, our contribution provides insight into the characteristics of non-verbal emotional expression among political leaders, as well as an open-source workflow for further computational studies of their affective communication.
\end{abstract}
\vspace{0.05cm}


\newpage

\section{Introduction}

In the growing body of literature on populism, the category of emotion has most frequently been utilized in discrete ways to describe its emergence and tenacity \citep{demertzis2006EmotionsPopulism2006}---by reference to the disillusioned silent majority, heightened emotions surrounding political leaders, and emotionally charged divisions between the people and the elite \citep{moffitt2014Rethinking,mudde2017PopulismIdeationalApproach2017a,norris2019CulturalBacklashTrump2019}. Compared to pluralist parties, who are commonly portrayed as more affectively neutral, populist rhetoric is most often characterized as deeply impassioned and imbued with strong (negative) emotions \citep{bonansinga2020WhoThinksFeels2020,salmelaEmotionalDynamicsRight2018}. Although the nature of the relationship between emotionality and populist discourse has rarely been subject to systematic empirical scrutiny \citep{widmannHowEmotionalAre2021}, this link has become one of the main focal points of research on political communication in recent years. Comprehensive research on the matter is thus crucial for gaining insight into the extent to which populist narratives and performances are rooted in and shaped by emotion, and by extension, how emotion may be utilized as a tool for building and maintaining political support \citep{nguyen_emotions_2019}.

In the past, studies of populist communication have generally focused on its verbal elements, by analyzing the affective content of the words used in oral or written addresses of political leaders \citep{bobba2019Social,breeze2019Emotion,breeze2020Angry,cochrane2022AutomaticAnalysisEmotion2022,jacobs2020TwitterFacebookPopulists2020,martellaPopulismEmotionsItalian2022,widmannHowEmotionalAre2021}. While advantageous in its own way, this type of analysis can pose a challenge to researchers primarily due to language barriers and the fact that written text usually contains only a small amount of words that have an inherent affective connotation \citep{widmannHowEmotionalAre2021}. At the same time, currently available sentiment analysis tools are predominantly based on and developed for the English language, while the output for other languages is often of lower quality \citep{zotero-18558}. These circumstances therefore limit the possibility of result generalization and cross-cultural comparison.

Visual analyses, although insufficiently represented in the given field of research, may provide further insight into the characteristics of affective populist communication as they are not hindered by the same limitations as text-based studies \citep{moffitt2022Taking} and allow for analysis of the visual rhetoric constructed and utilized by political leaders \citep{mendoncaPopulismParodyVisual2021}. Within the scope of visual research on political emotion display, two main schemes of analysis have shown to be especially valuable: the ethological approach \citep{masters1986FacialDisplaysLeaders1986} and the  Facial Action Coding System (FACS) \citep{ekman1976Measuring}.

Studies within the former largely focus on the analysis of political candidates' non-verbal representations in news coverage, their displays during political campaigns and debates, as well as the impact of different forms of representation on viewers and success in political elections \citep{bucy2020PerformingPopulismTrump2020,bucy2007TakingTelevisionSeriously2007,bucy2008HappyWarriorsRevisited2008,grabe2009ImageBitePolitics2009,sullivan1988HappyWarriorsLeaders1988}. The Facial Action Coding System, on the other hand, allows for a more in-depth analysis of the (micro)expressive behavior of political leaders during speeches and public performances. Due to its complexity, Ekman's classification system has been applied in a limited number of studies \citep{stewart2009TakingLeadersFace2009,stewart2015StrengtheningBondsConnecting2015,stewart2013Interpreting} but as the process of visual analysis gradually becomes automated \citep{joo2019Computational} this research method is becoming more accessible to researchers from various fields of interest \citep{ash2022Fight,bossetta2023CrossPlatform}.

Our study expands on the existing framework of FACS research on political emotion display and applies a computer vision approach for detecting emotions from the facial expressions of political leaders. Because it is relatively new, this approach faces various methodological challenges which we aim to address by following the recommendations outlined by \citet{barrett2019emotio}, such as using big data techniques and automated detection in naturalistic settings, integrating traditional methods with machine learning, and sampling visual data across diverse contexts and cultures.

Since the introduction of the FER dataset \citep{goodfellow2013Challenges}, different machine learning algorithms have been developed for solving the challenge of facial expression recognition, which essentially requires performing two tasks: (1) detection of facial expression, and (2) emotion recognition from the detected facial expressions. We develop a reproducible and principled computational workflow in Python for the analysis of video content of political leaders' public performances. Our work relies on the existing Python library \emph{fer} \citep{shenk2021JustinshenkFerZenodo2021} which bundles a pre-trained convolutional neural network (CNN) Keras models for face detection \citep[based on]{zhang2016JointFaceDetection2016} and emotion recognition \citep{arriaga2017Realtimea,arriaga2020Perception}.

We find that populist leaders on average express more negative emotions and that the emotionally neutral facial expression is less frequently detected in their public appearances compared to non-populist leaders. Overall, our work points towards a new direction in computational studies of the affective aspects of populist political communication and provides a roadmap for overcoming the limitations of the current approach in future studies.

\section{Populism as performance}

Defining populism represents a challenging task for researchers as there is barely any consensus on what such a definition should (or should not) include, and the concept itself becomes increasingly muddled as the term is used to describe a growing number of phenomena \citep{gidron2013VarietiesPopulismLiterature2013}. Most conceptualizations of populism can, however, be classified \citep{rovirakaltwasser2017OxfordHandbookPopulism2017} into one of three main approaches: (1) the ideational approach \citep{mudde2004PopulistZeitgeist2004}, (2) the discursive approach \citep{norris2019CulturalBacklashTrump2019,ostiguy2017PopulismSocioCulturalApproach2017} and (3) the political-strategic approach \citep{weyland2017PopulismPoliticalStrategicApproach2017}. For the purposes of this study, we adopt the discursive approach, which
conceives populism primarily as a rhetorical style based on first-order principles pertaining to \emph{who} should rule---the people, as ``the only legitimate source of political and moral authority in a democracy''---but which says very little about the \emph{how}  \citep{norris2019CulturalBacklashTrump2019}. Whereas the ideational approach claims populism is a ``thin-centered'' ideology \citep{mudde2004PopulistZeitgeist2004} and emphasizes the interpretation of politics as ``a Manichean struggle between a reified will of the people and a conspiring elite'' \citep{hawkins2018introduction} (Hawkins and Rovira Kaltwasser, 2018), the discursive approach highlights the ways in which this interpretation is framed and communicated as an anti-establishment discourse \citep{aslanidis2017avoidi}. We find this to be the most comprehensible understanding of populism as it considers the operationalized elements of the populist communication logic \citep{engesser_populist_2017} and examines the rhetorical components implicit in the ideological definition \citep{aslanidis2016populi}.

From this perspective, the populist rhetoric has widely been characterized by the use of ``dramatization, polarization, moralization, directness, ordinariness, colloquial and vulgar language'' \citep{engesser_populist_2017}, and low-order appeals within which leaders ``use a language that includes slang or folksy expressions and metaphors, are more demonstrative in their bodily or facial expressions as well as in their demeanor, and display more raw, culturally popular tastes'' \citep{ostiguy2017PopulismSocioCulturalApproach2017}. Because it represents a form of symbolic resistance to the more formalized and composed self-representation of mainstream pluralist leaders, i.e., the ``establishment'' and ``political elite'', the populist performance \footnote{By “the populist performance” we mean the populist communication style as laid out by \citet{bucy2020PerformingPopulismTrump2020}—that is, the verbal, tonal, and non-verbal display of a political leader which features three main communicative dimensions: simplification, emotionalization, and negativity.} is subversive to the normative communication structures of political dialogue. As such, it appeals to ``common sense'' \citep{bucy2020PerformingPopulismTrump2020} and claims to give a voice to ``the ordinary people'' \citep{inglehart2016TrumpBrexitRise2016}.

This mode of communication has shown to be highly effective as a means of garnering media attention and cultivating a dedicated base of supporters---especially in the new media environment---and populist leaders seem to be overall more adept at capitalizing on it than other, non-populist political figures \citep{bobba2019Social, davidson_are_2022}. Since digital media hinges on the attention economy \citep{wells_trump_2020} and thus tends to favor ``dark participation'' \citep{larsson_picture-perfect_2022}, populist leaders and political parties may fare well online due to the transgressive nature of their self-representation and the simplification, emotionalization, and negativity inherent in populist messaging \citep{bucy2020PerformingPopulismTrump2020,engesser_populist_2017}. Social media is a particularly convenient tool for the proliferation of populist messages and imagery as it allows political actors to circumvent the mainstream media, reach large audiences, and communicate and interact directly with `the people' \citep{lalancette2019Power,moffitt2022How}. At the same time, online platforms allow for an instant audience response, as the ``feedback loop between producer and consumer is much more intense and rapid than with television and radio'' \citep{finlayson_youtube_2022}. This form of `mediatized populism' \citep{mazzoleni2014Mediatization} allows for aspecific type of political performance, one which is largely unmitigated and relies heavily on portraying political leaders favorably to (online) audiences by framing them as expressive, dominant, and disruptive figures who are in tune with ``the will of the people''.

In recent years, much attention has been paid to the strategies of (affective) populist communication online, especially through analyses of verbal content shared by populist leaders and parties on Twitter and Facebook \citep[e.g.][]{bracciale2017Define,grundl2022Populist,jacobs2020TwitterFacebookPopulists2020,martellaPopulismEmotionsItalian2022,widmannHowEmotionalAre2021}. Given the rise of visual communication in politics and the rapid developments of Web 2.0 practices \citep{grabe2009ImageBitePolitics2009,larsson_picture-perfect_2022,messaris_digital_2019}, our primary aim was to expand the scope of research on the affective self-representation of populist leaders online by examining the characteristics of their non-verbal emotional communication. The facial expressions of political leaders are a particularly valuable area of research in this context, since digital media provide a breadth of material for analysis---well-lit, high-quality close-up shots of political leaders are now more prevalent than ever. We used this abundance of available data sources to our advantage and processed 220 YouTube videos uploaded to official accounts of political parties and their leaders, amounting to a total of 77 hours of video material. Our analysis examines potential differences in facial expressions between populist and non-populist political leaders during public performances, and more specifically, the characteristics of their negative emotional expressions and neutrality in emotional stance (i.e., absence of emotional expression).

\section{Data \& Method}

\subsection{Party selection}

To create our sample, we first had to differentiate and categorize political parties (and, by extension, their leading figures) as either populist or pluralist. For this purpose, we used data from the Global Party Survey (GPS) \citep{norris2020MeasuringPopulismWorldwide2020}, an expert survey from 2019 which offers an overview of key ideological and issue positions for 1043 parties from 163 countries worldwide. While expert surveys certainly have their limitations \citep{wiesehomeier2018expert}, we believe this database remains the most comprehensive comparative overview of political parties globally.

Among other data, such as ideological framework and party size, the survey also offers an estimate of the degree to which existing parties employ populist rhetoric. In the questionnaire design, populist rhetoric is defined as ``language which typically challenges the legitimacy of established political institutions and emphasizes that the will of the people should prevail'' \citep{norris2020MeasuringPopulismWorldwide2020}, whereas pluralist rhetoric ``rejects these ideas, believing that elected leaders should govern, constrained by minority rights, bargaining and compromise, as well as checks and balances on executive power'' \citep{norris2020MeasuringPopulismWorldwide2020}. The populist rhetoric scale (V8 in GPS
dataset) is based on expert assessments of rhetorical style given these parameters. The variable \textit{Type\_Populism} categorizes this scale into four ordinal groups: strongly pluralist, moderately pluralist, moderately populist, and strongly populist. Although these categories may not be able to fully capture the nuances of populist rhetoric---its culturally specific manifestations, or what ``the will of the people'' may connote for people across different political contexts---they allow insight into the fundamental characteristics of party rhetoric and enable fair comparison between countries.

Given that a great concern in expert surveys is the possibility of experts' backgrounds introducing cognitive bias into their assessments \citep{curini2010expert}, it is also important to note that internal validity testing done within the GPS showed that the personal characteristics of experts (such as ideological stance, gender, and nationality) were not significant predictors when it comes to the placement of parties on the Pluralism-Populism scale \citep{norris2020MeasuringPopulismWorldwide2020}.\footnote{For external validity measures, see robustness tests in \citet{norris2020MeasuringPopulismWorldwide2020}.}  Additionally, to establish that differences in expression of emotion are indeed more of a matter of populist leaning rather than ideological preference, we  use the ``Type\_Values'' variable in the supplementary analysis to examine ideological differences in emotion expression (see the Supplement section E -- Ideological differences).

From the GPS dataset, fifteen countries with competitive popular
elections were chosen: Argentina, Australia, Brazil, Croatia, France,
Hungary, India, Italy, Japan, the Netherlands, Serbia, South Africa,
Turkey, the United Kingdom, and the United States.\footnote{The average number of experts per country in the selected dataset was approximately 27, with the highest number being in the United States (50) and the lowest in Serbia (6). While this number is higher than the average for the GPS as a whole---19 \citep{norris2020MeasuringPopulismWorldwide2020}---the variance between countries could potentially entail variance in overall quality of assessment.} This sample, although limited, provides somewhat broad coverage of various cultures and political systems globally. Three parties with varying degrees of populist rhetoric---strongly populist, strongly pluralist, and either moderately populist or moderately pluralist---were selected from each country. We primarily focused on minor and major political parties as defined by the GPS variable categorizing parties based on their share of the vote in contests for the lower house of the national parliament. Fringe parties (less than 3\% of the vote) were generally excluded from the sample since their audience reach and influence is considerably less than that of their larger counterparts competing at the same scale in their respective countries. Likewise, due to this criteria, only two parties from the US were included in the sample. Table from the Supplement D -- Political leaders contains information about the selected parties and their leaders.

\subsection{Video selection}

Once party selection was complete, we identified one leader per party who has held a prominent
political position over the past decade (up to 2019, covered by the
GPS), and found their official YouTube channel or their party's official YouTube channel. From these channels, five videos were selected for each leader based on the following criteria:

\begin{enumerate}
\def\labelenumi{(\arabic{enumi})}
\item \emph{View count}. Videos that reached a larger audience were given priority over videos that did not perform comparatively as well on the  platform. View count was used as a filtering mechanism to single out videos with large and potentially diverse audiences, because the most viewed videos on the official YouTube channels of major political parties usually refer to topics of wider public interest. This is partially due to the ``popularity bias'' on YouTube, meaning it has a tendency to recommend political content that aligns with the interests of the majority \citep{Heuer2021-he}.
\item
  \emph{Unmediated representation}. Only direct and unmediated
  representations of the given political leaders were included, i.e.,
  videos which were not in any way previously modified by third parties
  such as news outlets. These consisted mainly of speeches, press
  conferences, ads, and promotional videos. This was done to ensure that
  the given results would reflect the leaders' self-representation,
  rather than their representation in news coverage.
\item
  \emph{Consistent framing}. A preference was given to videos where the
  face of the leader was in focus for an extended period of time and
  which did not frequently switch between disparate camera angles. This
  allowed for an overall consistent frame of reference for each given
  video.
\item
  \emph{Video quality}. High-quality, well-lit videos were selected
  whenever possible so as to improve analysis accuracy.
\end{enumerate}

This selection process resulted in a sample of 220 videos, representing 44 political leaders from 15 different countries. The sample is not gender-balanced as only five female leaders are present, which reflects the global underrepresentation of women in leadership positions within major political parties. As a result, our analysis does not account for gender, which we acknowledge as a limitation of our approach. Additionally, because the sample is limited by the timeframe of the currently available GPS dataset, it hasn't captured more recent global developments on the populist-pluralist spectrum (e.g. Giorgia Meloni and the Brothers of Italy movement or the Servant of the People party in Ukraine are not present in the dataset). Given these limitations, our sample is conceived as a diverse snapshot of the state of populist and pluralist rhetoric in the second decade of the 21st century and does not capture the full scope or evolution of populism today.

\subsection{Video processing and analysis}

Each video in the sample was processed by extracting 300 frames uniformly distributed over the length of the video, which were the basis for our main analysis. To improve the robustness of the overall approach, we also included an alternative frame processing approach in the supplementary material, based on processing every 50th frame of the video. Given that the majority of the YouTube videos have 24--30 frames per second, this is equivalent to selecting a frame every 1.67--2.00 seconds, which is a reasonable time window if we expect to capture potential changes in facial expression. The average number of frames per video extracted using this approach was 273.03, which makes the resulting time-series of emotion scores comparable in length with the first approach.

For each selected frame of every video in the sample the following procedure is applied. First, face detection is applied to the extract frames using  a three-stage cascaded CNN architecture with default parameters (scale factor: 0.709, minimum face size: 40 pixels, thresholds: [0.6, 0.7, 0.7] for the three detection stages) \citep{zhang2016JointFaceDetection2016}. This approach is particularly suitable for detecting faces in videos because it performs simultaneous face detection and alignment, which is important as we cannot expect that political leaders will be looking straight at the camera in each extracted video frame. After face detection was applied to all frames, there were videos in which multiple faces were detected throughout the frame (when a second face appeared in >5\% of frames). These videos were manually checked to confirm which of the detected faces corresponds to the selected political leader, and in some cases, specific frame ranges were specified to use only portions with consistent face detection.

The second step is emotion detection within the verified detected face box. We relied on implementing the mini-Xception architecture of CNN \citep{arriaga2017Realtimea} aimed at detecting 6 basic facial expressions of emotion, as well as a neutral expression. This model uses 4 residual depth-wise separable convolutions with input shape of 64x64 pixels (grayscale) and outputs 7 emotion scores. We initialized the detector with MTCNN enabled using the \texttt{FER(mtcnn=True)} constructor from the \emph{fer} library. Similar to the face detection approach outlined above, this implementation is aimed at real-time emotion detection from video input for robotic applications. This model achieves 66\% accuracy \citep[see][p.~3 for more details]{arriaga2017Realtimea} on the FER2013 dataset \citep{goodfellow2013Challenges} which is a standard benchmark for emotion detection. This dataset consists of 30,000 small images
(48x48 pixels) showing diverse facial expressions of different emotions. \footnote{FER2013 was constructed using Google Images API queries for queries for emotion-related keywords like ``blissful'' and``enraged'', combined with queries introducing diversity in gender, age and ethnicity represented in the images. Images were then labeled  by human coders \citep[pp.~3-4]{goodfellow2013Challenges}. Unfortunately, no information is provided by the authors of the dataset regarding the number of human coders nor their background, or the labeling methodology applied to extract training labels for every image
included in the FER2013 dataset. Benchmarks based on the FER2013 dataset should be taken with caution, especially in social science applications, because the dataset construction and validation documentation lacks crucial information about inter-coder reliability, coding procedures, and potential biases in the image selection process} Each emotional expression has approximately 6000 images, except disgust, which is only present in 600 images. According to \citet{goodfellow2013Challenges}, human classification accuracy is in the range of $65\% \pm 5\%$, but no additional information about the sample size, number of coders, or detection procedure is given.

In order to provide additional information and to validate the output of the machine learning approach, we conducted a small image annotation study where 5 coders annotated a sample of 514 randomly selected frames extracted from 220 videos. Coders received basic training in emotion recognition and, due to limited resources, answered only two questions about every image: (1) Can you clearly see the face of a politician in this image? (Yes/No), (2) Is this face expressing positive emotion, negative emotion, or is it a neutral expression? (``Positive emotion'', ``Negative emotion'', ``Neutral expression''). The majority vote label was set as the ``human-coded label'' and compared with the output of the machine learning algorithm, whose label is defined based on the highest score out of the sum of positive emotions, negative emotions, or the neutrality score. A detailed description and the results of the data annotation study can be found in Supplement F -- Image annotation.

The result of the emotion detection procedure on a single video frame is an array of 7 emotion scores, each representing the predicted proportion of an expressed emotion. Each score is matched with one of 7 labels: ``angry'', ``disgust'', ``fear'', ``happy'', ``sad'', ``surprise'' and ``neutral''. The lowest possible score indicating the absence of a particular emotion is 0, while the highest score, which indicates a ``clean'' expression of a single emotion, is 1. In a statistical sense, the resulting data are compositional as they do not carry absolute, but rather relative information about the expressed emotions. This data is organized as a data frame consisting of 7 columns and approximately 300 rows, storing emotional scores for each selected frame of the video.

\begin{figure}[ht!]
     \centering
     \includegraphics[width=0.48\textwidth]{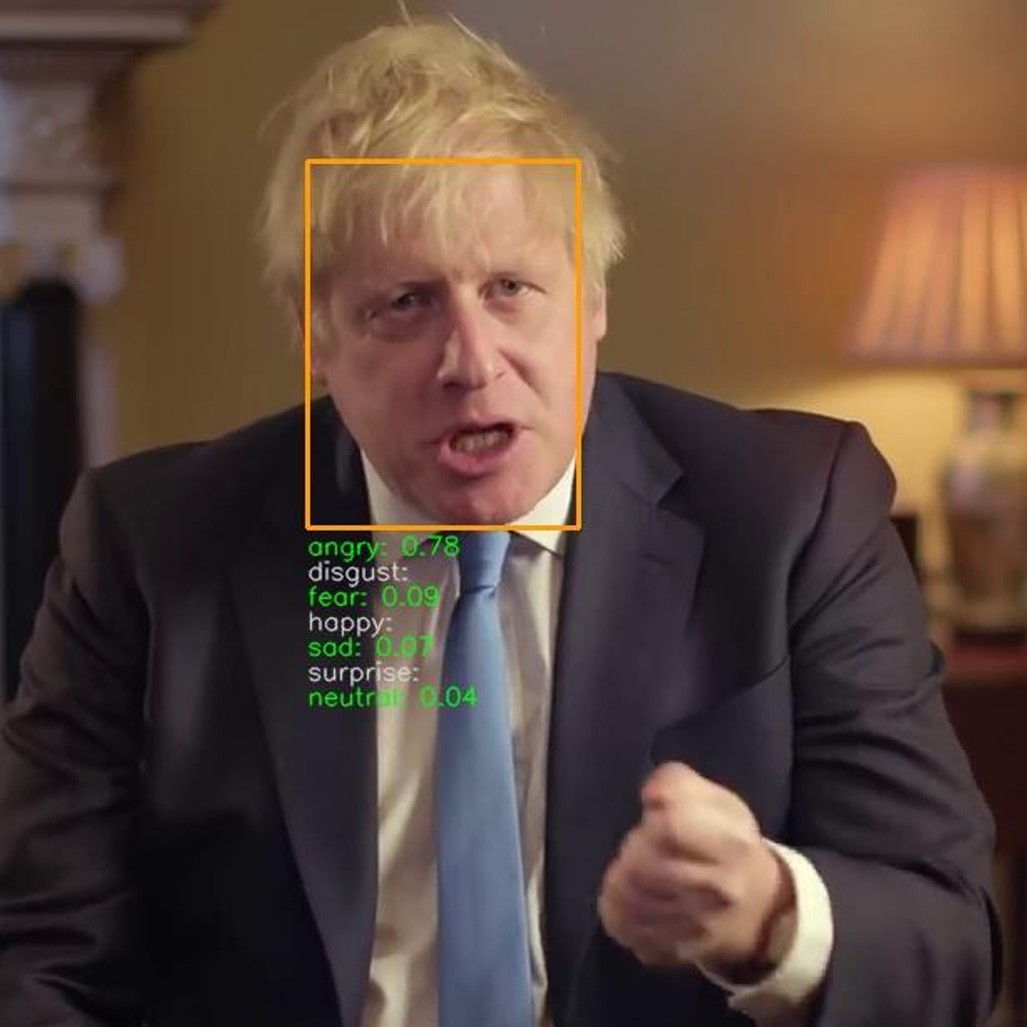}
     \includegraphics[width=0.48\textwidth]{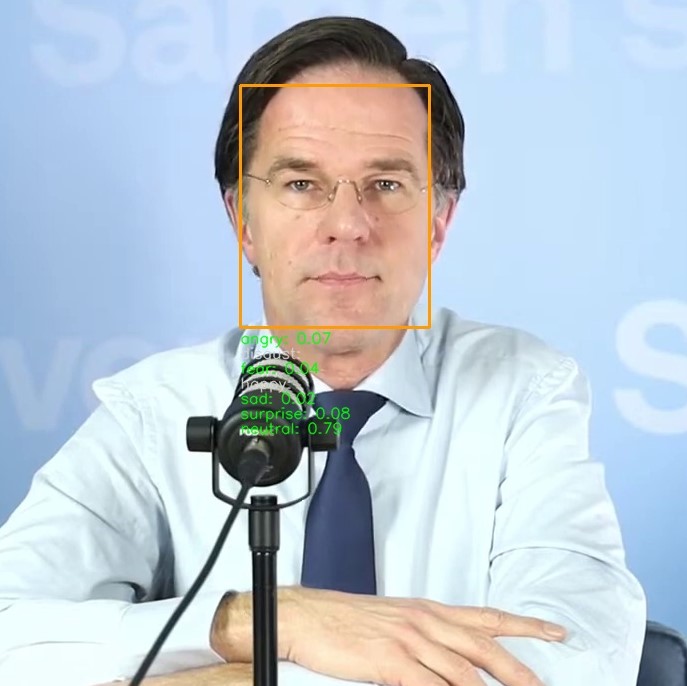}
     \caption{Example frames showing a high detected score of negative emotions (left: angry 0.78, disgust 0.00, fear 0.09, happy 0.00, sad 0.07, surprise 0.00, neutral 0.04) and a high detected score of neutral expression (right: angry 0.07, disgust 0.00, fear 0.04, happy 0.00, sad 0.02, surprise 0.08, neutral 0.79).}
     \label{fig:emo-ex}
\end{figure}

Because we were primarily interested in neutral emotional expressions and expressions of negative emotions, we created an additional variable
which sums the scores of negative emotions: anger, fear, disgust, and sadness. Figure \ref{fig:emo-ex} shows examples of faces with a high detected score of negative emotion and a high score of neutral expression.

The entire analysis was performed in Python and the code is publicly available in the GitHub repository: \href{https://github.com/atomashevic/face-of-populism}{https://github.com/atomashevic/face-of-populism}. The code is organized into three sections. The data processing part of the code can be used to reproduce emotion score data files from the list of YouTube URLs for the videos included in the sample. It can easily be modified to expand the sample to a larger number of videos. Face and emotion detection relies heavily on the \emph{fer} Python library and that part of the code is presented in the Jupyter notebook. Finally, we have data analysis scripts for performing data wrangling and statistical tests as well as producing Figures \ref{fig:pop2}, \ref{fig:rain-neg}, and \ref{fig:rain-neu}.

\subsection{Data processing pipeline}

Our complete data processing pipeline consists of several sequential steps:

\begin{enumerate}
\item \textbf{Video acquisition:} Videos were downloaded from YouTube using the \emph{pytube} library.
\item \textbf{Frame extraction:} Using the OpenCV (\emph{cv2}) library, we extracted video metadata (fps, frame count, duration) and selected frames according to our sampling method.
\item \textbf{Face detection:} The MTCNN model implemented in the \emph{fer} library detected and aligned faces in each frame.
\item \textbf{Face verification:} Videos with multiple detected faces were manually reviewed to identify the political leader.
\item \textbf{Emotion recognition:} The mini-Xception CNN model processed each detected face to output 7 emotion scores.
\item \textbf{Data aggregation:} For each video, we calculated mean scores for each emotion category and the composite negative emotion score.
\item \textbf{Statistical analysis:} We performed t-tests, ANOVA with post-hoc Tukey HSD tests, and calculated effect sizes to examine differences between populist and pluralist leaders.
\end{enumerate}

All code for this pipeline is organized in a publicly available repository with a structured organization where raw data is processed through these steps, with intermediate and final results stored in appropriate directories. Videos requiring specific handling (e.g., cropping to specific time ranges) were processed using the \emph{moviepy} library before entering the main pipeline.

\section{Results}

In the entire sample, there are 203 successfully processed videos using the approach of selecting 300 frames uniformly distributed across the entire length of the video clip. For 17 videos, multiple faces were detected without consistent classification of each person as a unique face, and these videos were discarded from the analysis. Pluralist leaders are present in 96 videos (31 strongly pluralist and 65 moderately pluralist) and populist leaders are present in the remaining 107 videos (69 strongly populist and 38 moderately populist). For every video, the emotion recognition procedure returns the score of the 6 detected emotions (anger, disgust, fear, happiness, sadness, surprise) as well as the score of the neutral expression for each of its frames. These scores sum up to 1 and they can be viewed as proportions of each emotion present in the compound facial expression detected in the frame.

Human image annotation was performed on a sample of 514 images extracted from 220 videos. Agreement between human coders ranged from 30.3\% to 69.8\%, while Cohen\textquotesingle s Kappa coefficient varied from 0.15 to 0.462. Overall, Fleiss\textquotesingle{} Kappa is 0.231, which suggests fair agreement between coders but is also indicative of how difficult the task of recognizing the emotional expression in a single, randomly selected video frame is. When only selecting cases where a majority vote label exists (458 images), agreement between that label and the result of the machine learning classification is 55.2\%, with a Cohen's Kappa of 0.311 and F1 score of 0.502. This suggests moderate agreement and performance of the machine learning model relative to human consensus. When only selecting cases where the majority of coders and the model agree that a face is clearly visible (present) in the image (356 cases), agreement is 59.8\% (Kappa 0.320, F1 score 0.530).

For each video, we calculated the mean score of each emotional state for all frames, as well as the mean of neutral expression, and the mean of negative emotional expression for the entire video. Descriptive statistics of these values are presented in Table \ref{tab:desc}.

\begin{table}[ht!]
    \centering
    \begin{tabular}{|l|c|c|c|c|c|}
        \hline & Mean & SD & Min. & Median & Max. \\
        \hline Anger & 0.227 & 0.134 & 0.034 & 0.200 & 0.726 \\
        \hline Disgust & 0.003 & 0.008 & 0.000 & 0.001 & 0.066 \\
        \hline Fear & 0.127 & 0.091 & 0.014 & 0.099 & 0.548 \\
        \hline Happiness & 0.077 & 0.083 & 0.000 & 0.051 & 0.414 \\
        \hline Sadness & 0.206 & 0.110 & 0.023 & 0.181 & 0.568 \\
        \hline Surprise & 0.031 & 0.037 & 0.000 & 0.020 & 0.273 \\
        \hline Neutral & 0.328 & 0.169 & 0.020 & 0.300 & 0.807 \\
        \hline Negative emotions & 0.563 & 0.185 & 0.156 & 0.570 & 0.964 \\
        \hline
    \end{tabular}
    \caption{Descriptive statistics for extracted emotions for the entire sample, $N=203$ videos}
    \label{tab:desc}
\end{table}

Results suggest that negative emotions are more frequently expressed in
the entire sample, with anger (0.227) as the emotion having the highest
average score, followed by sadness (0.206) and fear (0.127). As a
consequence, the average proportion of negative emotions is greater than 0.5, which means that on average negative emotions are detected more frequently than positive emotions (e.g., happiness) and neutral
expressions combined. There are clear issues with the detection of
disgust and surprise, where a low maximum level suggests low positive
predictive value for these emotions. However, our focus is on
differences in the expression of negative emotions and in scores of
neutral states. Aggregating negative emotions enables us to surpass the
false positive issues within the group of negative emotions \citep[e.g.,
expression of fear being classified as sadness, disgust as anger,][]{khaireddin2021FacialEmotionRecognition2021} and minimizes the impact of the low positive predictive value of disgust and surprise.

\begin{figure}[ht!]
    \centering
    \includegraphics[width=0.75\textwidth]{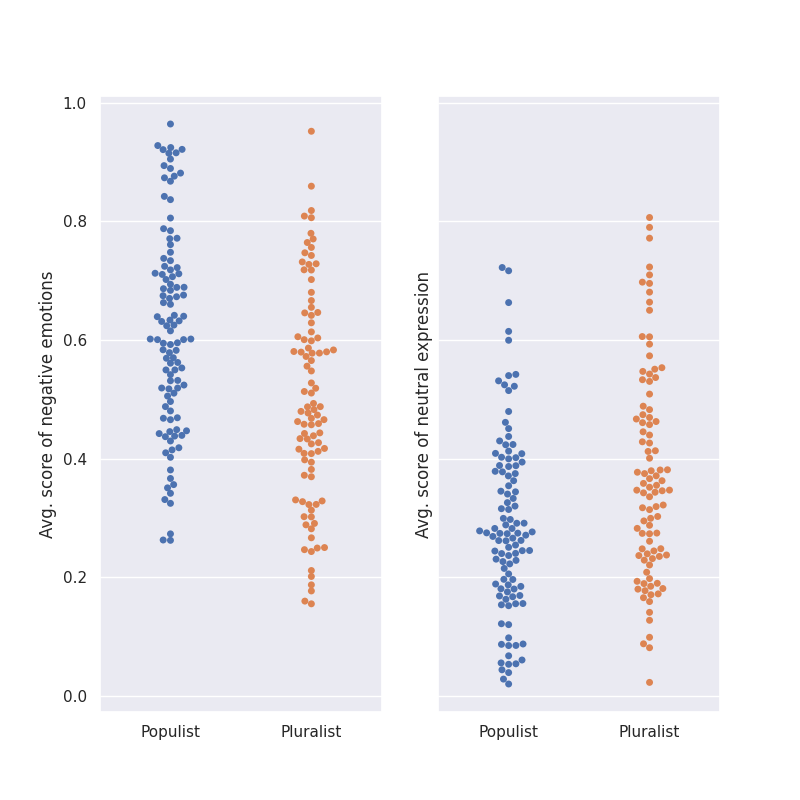}
    \caption{Distribution of average scores of negative emotions (left) and neutral expression (right) of videos featuring populist and pluralist leaders.}
    \label{fig:pop2}
\end{figure}

Looking at the maximum values of average detected scores, the largest
average scores of specific negative emotions were found in videos of
populist leaders. The highest proportion of anger was detected in a
video of Andrej Plenković (HDZ, Croatia), disgust in a video of Narendra Modi (BJP, India), fear in a video of Saša Radulović (DJB, Serbia) and sadness in a video of Donald Trump (GOP, USA). The video with the highest aggregate proportion of negative emotions, where one of the negative emotions is dominant in almost every analyzed frame, is the video titled ``Special Message from President Trump'', published on December 23, 2020. On the other hand, the highest average score of neutral expression is found in a video of Yukio Edano (CDP, Japan), whose party belongs, in contrast with previous examples, to the moderately pluralist group.\footnote{This does not imply that, on
  average, there is a tendency for Japanese leaders to be more neutral.   In fact, in Japan, we observe the same tendency as seen in the entire sample---the speeches of Shinzo Abe (LDP) from the populist group have the highest average score of negative emotions (0.58), while Natsuo Yamaguchi (NKP) from the moderately pluralist group has the lowest average score (0.30). Conversely, the average scores of neutrality show the opposite trend. These results are not extreme compared to the
  sample distributions shown in Figure \ref{fig:pop2}.}

To further examine differences between pluralist and populist leaders,
we used the GPS measure of populist rhetoric in a binary way to split
the sample into groups of pluralist (\emph{Party Populism} 1-2) and
populist (\emph{Party Populism} 3-4) leaders. The distribution of the
average scores of negative emotion and neutral expression is shown in
Figure \ref{fig:pop2}.

Overall, all average values have a large spread in both groups, with
pluralist leaders having several videos with less than a 0.2 average
negative emotion score and more than a 0.75 average score of neutral
expression. In terms of central tendency, we see denser grouping of
pluralist leaders between the 0.4 and 0.6 negative emotions score, while
populists tend to cluster in areas between 0.6 and 0.8. In terms of
neutral expression, we have a thicker cluster of points for populists in
the range between 0.2 and 0.4. Figure 1 suggests that, for populist
leaders, we are more likely to observe lower average scores of neutral
expression and higher average scores of negative emotion. The reverse
tendency can be observed in the case of pluralist leaders. Negative
emotions are dominant in 73.8\% of populist videos, compared to 45.8\%
of pluralist videos.

The mean average score of negative emotions in the populist group is
0.616 (95\% CI 0.584--0.649), compared to the 0.500 mean average score
for pluralists (95\% CI 0.464--0.537). The difference between the two
groups is statistically significant ($t=4.691$, $p=0.000005$, $d=0.66$ moderate effect size). This finding
suggests that pluralists generally tend to have an average score of
negative emotions around 0.5, which roughly translates to the situation
where negative emotion is the prevailing expression in one half of the
video (in fact, that percentage is 46.1\% compared to 66.7\% for
populists; see table in the Supplement B -- Dominant emotions per frame). On the other hand, for
populists, more than one half of each video where they appear is
dominated by negative emotions.

\begin{figure}[ht!]
    \centering
    \includegraphics[width=0.6\textwidth]{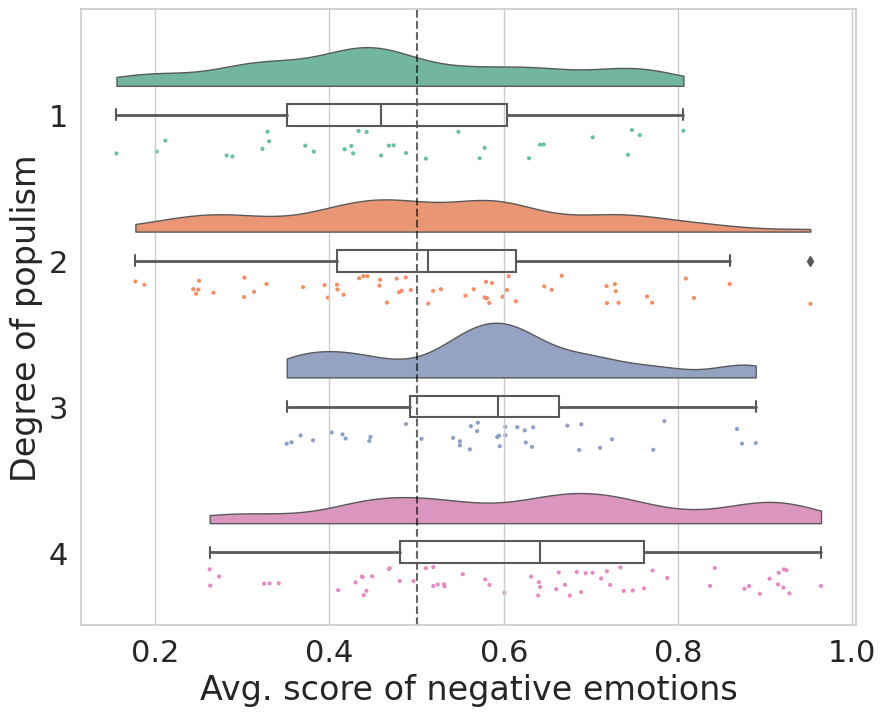}
    \caption{Raincloud plot showing the distribution of average
scores of negative emotions across four degrees of populism. Every data
point is presented as a colored dot and above them boxplots and density
plots are displayed. The dashed vertical bar indicates the score of
0.5.}
    \label{fig:rain-neg}
\end{figure}

The mean average score of neutral expression for populists is 0.289
(95\% CI 0.260--0.318), compared to the 0.374 mean for pluralists (95\%
CI 0.338--0.409). While this difference in means is statistically
significant ($t=-3.636$, $p=0.00035$, $d=-0.515$ moderate effect size), it is smaller than the one observed
in the case of negative emotions.

\begin{figure}[ht!]
    \centering
    \includegraphics[width=0.6\textwidth]{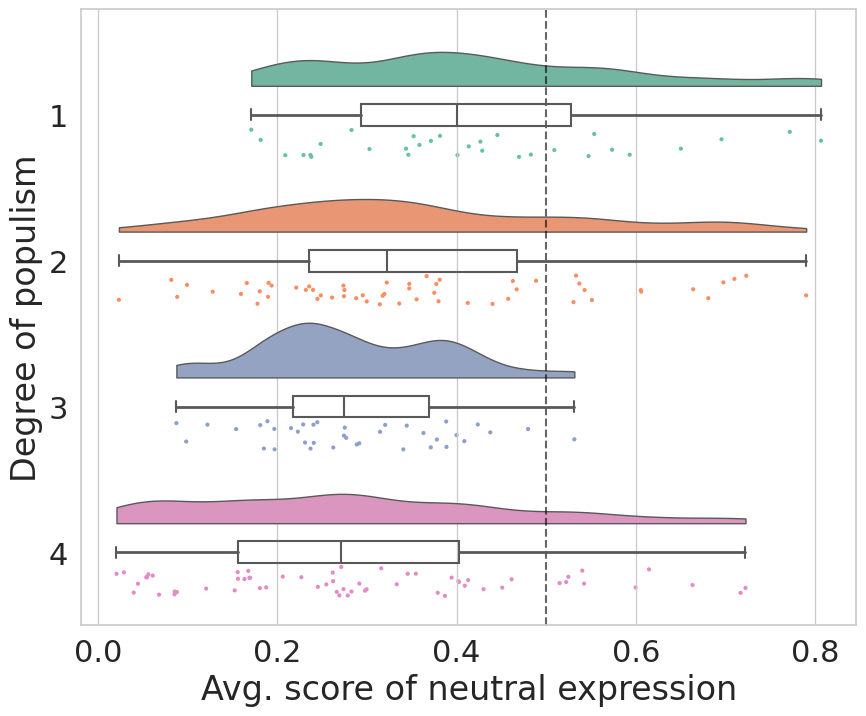}
    \caption{Raincloud plot showing the distribution of average
scores of neutral expression across four degrees of populism. Every data point is presented as a colored dot and above them boxplots and density plots are displayed. The dashed vertical bar indicates the score of 0.5.}
    \label{fig:rain-neu}
\end{figure}

To investigate the robustness of these differences, we took into account the differences between moderate and strong pluralists/populists provided by the GPS. The distributions of average scores in these four groups are compared and shown in Figures \ref{fig:rain-neg} and \ref{fig:rain-neu}.

In the case of negative emotions (Figure \ref{fig:rain-neg}), we observe four flat distributions with thick tails. However, the median point of all distributions progressively moves towards more negative emotions as we move from strong pluralists to strong populists. What is apparent in these differences is the low frequency of videos featuring populist leaders where the average score of negative emotions is below 0.5. It is worth pointing out that examples of extremely high scores of negative emotions are present also in pluralist groups.

Similarly, in the case of neutral expression (Figure \ref{fig:rain-neu}), we have flat distributions (except in the case of moderate populists) with long right tails containing videos of leaders who maintain a neutral expression throughout most of their recorded appearance. Again, median values move progressively towards lower neutrality as we move from pluralists towards populists, except that no large difference can be seen when comparing strong and moderate populists. In the case of strong populists, however, we have a larger number of videos where the neutral expression is almost completely absent.

Overall, the results of one-way ANOVA suggest that the differences
between the group means are statistically significant both in the case
of negative emotions ($F=7.109$ $p=0.000152$), with strong effect size
measured by eta-squared (0.104), and neutral expression ($F=5.625$
$p=0.001038$), with moderate effect size (0.084). Results from Tukey HSD
post-hoc tests reveal significant differences between leaders coming
from parties employing strong populist rhetoric and leaders coming from
pluralist parties (mean difference $0.1559$ $p = 0.0005$ in comparison with
strong pluralists, and mean difference $0.1148$ $p=0.0029$ in comparison
with moderate pluralists). For neutral expression we find the opposite
case, as strong pluralists differ from both groups of populists (mean
difference $0.135$, $p=0.0056$ for moderate populists, and mean difference
$0.129$ $p=0.0032$ for strong populists).

Overall, these results reveal differences in the average scores of
negative emotion and neutrality between populist and pluralist leaders
on a modest sample. These differences are more pronounced when we focus
on candidates coming from parties who strongly rely on either pluralist
or populist rhetoric, with strong pluralists on average favoring more
neutral expressions, while strong populists more frequently express
negative emotions---meaning that, for the majority of their public
appearance, we can observe facial expressions of negative emotions.

\section{Discussion}

The present study examines facial emotional expressions of populist and non-populist political leaders across fifteen countries using a deep-learning-based
computer-vision algorithm to investigate differences in their expressions of negative emotions (anger, disgust, fear, and sadness) and neutral stances, i.e., facial expressions which lack emotional disposition. Our findings indicate that populist leaders tend to express negative emotions more frequently than non-populist leaders, and conversely, that neutral expressions are less common in their public performances than they are in those of more conventional, pluralist political figures. The significance and the effect size of these differences are higher compared to differences between ideological groups (see Supplement E -- Ideological differences).
These findings align with the theoretical framework on the characteristics of affective populist communication \citep{bonansinga2020WhoThinksFeels2020,salmelaEmotionalDynamicsRight2018,engesser_populist_2017}, as well as
with previous, mostly text-based research which indicates that populist leaders employ language which is associated with negative emotion at a higher rate than mainstream, pluralist politicians \citep{breeze2019Emotion,caiani2023Populism,macagno2022Argumentation,martellaPopulismEmotionsItalian2022,widmannHowEmotionalAre2021}.

Overall, the results provide new evidence of a style of affective, non-verbal communication more common to political leaders whose rhetoric is populist-leaning, and thus further substantiate the idea that populism is performative in nature \cite{bucy2020PerformingPopulismTrump2020, moffitt2022Taking}. As such, because it is subversive in relation to established political conventions, the populist performance is especially well-suited for the environment of social media where the principle of ``any press is good press'' often applies---that is, the quantity of audience attention bears more weight than its quality \citep{wells_trump_2020}. On one hand, the populist use of negative emotional displays in videos may therefore be a mechanism through which they maintain relevance, considering that controversial content online ordinarily generates more engagement among consumers \citep{larsson_picture-perfect_2022}. On the other, populists' facial expressions can provide affective heuristics for audiences, influencing their evaluation of candidates' political competence \citep{boussalis2021FacingElectorateComputational2021} and allowing them to quickly reduce complex political issues into simple, emotional interpretations by visually framing them as negative.

Since this emerging dynamic has the potential to influence the behavior of voters---and perhaps even politicians themselves\footnote{Due to the unmediated relationship between political figure and voter, it may very well be possible for online audiences to directly influence politicians in return. For example, once videos prominently featuring negative emotions start performing well, they might create a feedback loop which, in the long run, incentivizes political figures to ``go negative" when they may have otherwise taken a different communication approach.}---it holds several implications for populism, political communication, and polarization research. The Manichaen mentality of ``Us versus Them" is fundamental to the populist worldview, and it further entrenches ideological positions and social divides. It is also largely dependent on fostering contentious discourse practices and negatively portraying political opponents and their beliefs. Thus, understanding how and when populist leaders bring negative emotional appeals to the forefront is vital for developing a more refined model of populist communication that would consider both the cognitive and affective sides to their political behavior. Studying non-verbal cues, in particular, can provide a breadth of information for this line of research since facial expressions of emotion add a layer of meaning to political messaging, establishing credibility and trust (or lack thereof) in political figures and appealing to audiences on a more intuitive perceptive level.

Apart from these findings, the present study introduces an open-source workflow for the analysis of political videos and provides a new open dataset of time-series emotion scores, as well as tools for the annotation of images in R and a new dataset of human labelled images which can be used to benchmark different machine learning approaches to facial expression recognition. Overall, these contributions aim to make large-scale computational studies of facial expressions of emotion more accessible and rigorous in the future.

Given the exploratory nature of the study and its reliance on machine learning algorithms for emotional detection, there are several notable limitations with significant impact on the results presented in this paper. There is a group of limitations inherent to the way we constructed the sample, and another related to the machine learning approach to detecting emotions from facial expressions.

Firstly, the sample of YouTube videos is not globally representative and does not provide ample coverage of the global political spectrum in regard to the use of populist or pluralistic rhetoric. As we have shown in the supplementary analysis (see Supplement C -- Country-level differences), the differences between populist and pluralist leaders are not consistent in all investigated countries. Larger research infrastructure is required to process a more substantial number of videos, leaders and countries and the current computational toolset needs to be extended to enable seamless automatic processing of videos. This also relates back to the gender imbalance of the current sample.
While it is reasonable to hypothesize that gender influences emotional expression among political leaders \citep{Boussalis2021-rb}, in this case, including more women without significantly increasing the overall size of the sample and the number of leaders would mean focusing only on countries where there is at least one female leader of a non-fringe political party. This creates the risk of introducing bias towards Western countries in the sample and disregarding the global scope.

Secondly, a key component of our analysis is the measure of populist rhetoric supplied by the Global Party Survey database \citep{norris2020MeasuringPopulismWorldwide2020} which significantly limits the timeframe of the analysis (no leaders,
parties and videos after 2019 were taken into consideration), but also imposes the inherent limits of an expert survey. While the GPS is the only available source of multi-dimensional measurements of populist rhetoric on a global scale, the experts surveyed may operate within different understandings of what populist rhetoric is. This is particularly challenging in developing countries, where fewer experts answered the survey, as well as in autocratic regimes where experts may not have been able to provide a fair assessment of the (ruling) political parties (or refused to do so). By focusing on large countries with free elections, we tried to mitigate some of the risks, but there is the possibility that some measurements used in this study reflect expert biases, especially since we used a single, aggregate measure of
populist rhetoric. Future studies should aim for more granularity and robustness in the measurement of populist rhetoric, which may be obtained by following a text-as-data approach and obtaining the measures using computational and machine learning tools, where expert survey scores provide external validation \citep{Jankowski2023-ft}.

Furthermore, since we lacked the resources necessary to process every political speech from each party's official YouTube channel, the first criterion of selection is the view count of the videos. Our expectation is that this metric serves as a proxy for the relevance of the speech content to a wider audience over a longer period of time. However, since we have no information about the audience, it is possible that we are observing a demand-side effect, where audiences who consume populist videos might show a stronger preference for those that prominently feature negative emotions, introducing a specific bias in the sample.

This consideration may apply even more broadly, as our analysis does not fundamentally account for systemic bias in media consumption, coverage, and narrative construction, and consequently the differences in emotional expression may not be a result of varying political styles but rather of bias in media selection. Future research should address this issue by comparing emotional expressions across a wider range of media formats---including other social media platforms, television broadcasts of speeches and debates, mediated interviews, etc.---to determine whether the observed patterns persist beyond YouTube's algorithm and environment.


The next major limitation is a result of the machine learning model deployed for emotion recognition. We conducted a small validation study to compare the performance of our model with the consensus of 5 human coders. Although results show evidence of moderate agreement between humans and the machine learning model (53-60\%), there is high inconsistency between human coders even for the simple task of assessing whether a given facial expression shows positive emotion, negative emotion, or a neutral expression. This level of agreement, while adequate for our exploratory study, indicates substantial uncertainty in the classification of emotional expressions that could potentially affect our conclusions. The higher detection rate for negative emotions compared to positive ones could potentially amplify differences between groups if the model is more sensitive to certain types of expressions.

There is an urgent need for large-scale image annotation studies with expert coders if we are to assess the performance and accuracy of FER models in the future, especially on difficult ``in the wild'' image datasets such as ours. However, it is worth noting that in our main analysis we focus on averages across 300 frames of each video, making our analysis robust to misclassification of a smaller number of individual frames. This methodological choice helps mitigate the impact of individual frame misclassifications on our overall findings. That being said, we believe that fair agreement between the human majority vote and machine learning model does not compromise the validity of our findings.

For future research, several technical improvements could enhance the accuracy of facial emotion recognition in political contexts. The most promising direction appears to be zero-shot classification based on open-source transformer models \citep{min2024emotio} such as OpenAI's CLIP (Contrastive Language-Image Pre-training) \citep{radford2021learni}, which may offer more nuanced performance with the ability to process facial expressions in the specific context of the image along with other visual cues. These models can leverage large-scale pre-training on diverse datasets and potentially provide more culturally sensitive interpretations of facial expressions \citep{waligora2024joint}. Additionally, fine-tuning pre-trained models specifically on datasets of political figures' expressions could improve domain-specific accuracy, as could ensemble methods that combine multiple classification approaches to enhance robustness.

Overall confidence in the presented results would surely be higher if we based our analysis on a computer-vision approach which was shown to have higher accuracy, at least on FER2013 data. One of the possible alternatives would be to follow recent work \citep{bossetta2023CrossPlatform,schmokel2022FBAdLibrarian} and use proprietary API such as Amazon's Rekognition. Nonetheless, the aim of this paper was to introduce an open-source approach focusing on free computing resources and the deployment of machine learning algorithms on systems that are accessible to individual researchers in social sciences, including graduate students without significant institutional support for computational work.

This study was focused on exploring basic mean-level differences between the given groups, but it is worth noting that the full output of our emotion detection workflow includes time series data suitable for more advanced analyses. For instance, multivariate analysis could explore affective dynamics \citep{ryan2024toward}, identify differences related to various factors (such as type of public address or proximity of elections), and cluster political leaders according to temporal patterns of emotional expression. Furthermore, multivariate change point detection methods \citep{james2015ecp} could identify key moments where emotional expressions significantly change. Future research could extend this approach in other promising directions as well. Network psychometric methods could model dynamic interrelationships between emotional expressions, potentially revealing different temporal patterns between populist and pluralist leaders \citep{tomasevic2024dynami}. Recent speech-to-text models like Whisper \citep{radford2022robust} could enable integrated analyses of verbal and non-verbal emotional expressions, providing insights into the alignment between spoken content and facial displays. Additionally, simulation approaches could help test theoretical predictions about emotion dynamics in political communication \citep{tomasevic2024decodi}. Collectively, these approaches would move beyond static comparisons toward analyses of temporal dynamics, potentially revealing how leaders modulate emotional displays in response to audience feedback or political contexts.

As computational studies of political communication rapidly embrace machine learning tools and algorithms, this study offers an extension of an open-source toolbox for the analysis of affective non-verbal political communication. Analyses presented in this paper are based on a novel dataset which is only a starting point for further studies, and which will hopefully expand in the future to encompass a more diverse sample of political leaders in videos obtained from different platforms. Gaining this kind of insight into the visual aspects of emotional political communication, coupled with the existing analyses of its verbal elements, holds the potential for a more comprehensive understanding of how leaders convey political messages, persuade audiences, and present themselves to the public using digital media platforms.

\section{Data availability}

Data files are stored as CSV files. Processed
data used for both the main and supplementary analyses can be found at:
\href{https://github.com/atomashevic/face-of-populism}{https://github.com/atomashevic/face-of-populism}.
Repository contains a CSV file with a list of URLs of all YouTube videos used in the analysis as well as Python code that can be used to reproduce the CSV files with processed data. Code used for image annotation and the resulting dataset can be found at: \href{https://github.com/atomashevic/fop-annotation}{https://github.com/atomashevic/fop-annotation}.

\section{Acknowledgements}\label{acknowledgements}

The authors did not receive any financial support for this research. The annotation of images was made possible thanks to the dedicated work and effort of our graduate students, A. Na\dj{}, M. Milanko, and A. Mi\v{s}i\'{c}, who volunteered as coders.

\bibliography{fop}
\end{document}